# Observation of Giant Quantized Phonon Modes in Graphene via Tunneling Spectra

Yu Zhang, Qian Yang, and Lin He[*]

Center for Advanced Quantum Studies, Department of Physics, Beijing Normal University, Beijing, 100875, People's Republic of China.

[*]Correspondence and requests for materials should be addressed to L.H. (e-mail: helin@bnu.edu.cn).

**Phonons, the fundamental vibrational modes of a crystal lattice, play a crucial role in determining electronic properties of materials through electron-phonon interaction. However, it has proved difficult to directly probe the phonon modes of materials in electrical measurements. Here, we report the observation of giant quantized phonon peaks of the *K* and *K'* out-of-plane phonon in graphene monolayer in magnetic fields via tunneling spectra, which are usually used to measure local electronic properties of materials. A perpendicular magnetic field quantizes massless Dirac fermions in graphene into discrete Landau levels (LLs). We demonstrate that emission or absorption of phonons of quasiparticles in the LLs of graphene generates a new sequence of discrete states: the quantized phonon modes. In our tunneling spectra, the intensity of the observed phonon peaks is about 50 times larger than that of the LLs because that the *K* and *K'* out-of-plane phonon opens an inelastic tunneling channel. We also show that it is possible to switch on/off the quantized phonon modes at nanoscale by controlling interactions between graphene and the supporting substrate.**

The electronic properties of materials can be strongly modified by phonons through electron-phonon interactions [1]. The most famous phenomenon induced by electron-phonon coupling is the emergence of superconductivity in Bardeen-Cooper-Schrieffer (BCS) superconductors [2,3]. However, it is difficult to directly detect the phonon modes in electrical measurements. This is especially the case in graphene because the strength of electron-phonon interactions in intrinsic graphene is usually negligible due to the extremely weak electron-phonon pairing potential and a vanishing density of states (DOS) near the Dirac point [4-18]. Among all the phonons in graphene, $K$ and $K'$ out-of-plane phonon really stands out: it introduces a new inelastic tunneling channel in the perpendicular direction of graphene and generates a gap-like feature (with the "gap" of about 130 meV) exactly at the Fermi level ($E_F$). Such a characteristic feature has been extensively observed in the studies of graphene by using both scanning tunneling spectroscopy (STS) [19-27] and angle-resolved photoemission spectroscopy (ARPES) [28-30], which indicates that it is possible to detect some particular phonon modes of graphene in electrical measurements. In this Letter, we report the observation of giant quantized phonon peaks of the $K$ and $K'$ out-of-plane phonon in graphene monolayer under magnetic fields via STS spectra. Emission or absorption of the phonons of quasiparticles in the Landau levels (LLs) of graphene results in the emergence of giant quantized phonon modes, with the signal that is about 50 times larger than that of the LLs in our experiment. We also demonstrate the ability to switch on/off the giant quantized phonon modes at nanoscale by changing interactions between graphene and substrate.

The graphene monolayer is directly synthesized on Cu foils via a traditional low-pressure chemical vapor deposition (LPCVD) method, and then the transfer-assisted method is adopted to transfer the graphene sheet onto a 300-nm $SiO_2$/Si wafers [6,31,32]. In our experiment, we transfer three layers of graphene on $SiO_2$/Si wafers and the underlying two graphene sheets are used to reduce any possible interactions between the topmost graphene sheet and the substrate (the growth process and the characterization of the layer number are shown in Fig. S1 and S2 of the Supplemental Material [33]). Our experiment indicates that this layer-by-layer transfer process can effectively reduce the interlayer coupling, leaving the topmost graphene sheet behaves as a pristine graphene monolayer. Figure 1(a) shows a representative scanning tunneling microscopy (STM) image of the decoupled graphene monolayer. Due to the roughness of the underlying $SiO_2$/Si wafers, there exist surface corrugations with a vertical dimension of ~80 pm, leaving some nanoscale graphene regions suspended. Figure 1(b) shows a typical *dI/dV* spectrum recorded on a suspended graphene region, exhibiting a gap-like feature of approximately 130 meV pinned to $E_F$. Such a gap-like feature, which has been demonstrated explicitly in previous STM [19-27], inelastic x-ray scattering [34], and ARPES experiments [28-30], is attributed to the phonon-mediated inelastic tunneling. We can directly deduce the energy of the *K* and *K'* out-of-plane phonons, ~ 65 meV, from the corresponding $d^2I/dV^2$ spectrum, as shown in inset of the Fig. 1(b).

Figure 1(c) shows a representative *dI/dV* spectrum recorded on the suspended graphene region in the magnetic field of $B = 10$ T. Strikingly, the spectrum exhibits a series of pronounced peaks, which are almost symmetric about $E_F$ and show different features from that of the LLs in graphene monolayer. To explore the exact nature of these peaks, we carry out the STS measurements under different magnetic fields. Figure 1(d) shows the evolution of the *dI/dV* spectra as a function of magnetic fields *B* from 4 T to 12 T as a variation of 0.5 T. Each panel of every magnetic field consists of thirty *dI/dV* spectra acquired at different spatial points along a line of 20 nm. The spectra recorded at different positions are reproducible, indicating that the peaks in the spectra are only controlled by the magnetic fields. A notable feature is that there are two magnetic-field-independent peaks on both sides of the $E_F$ ($P_{0+}$ and $P_{0-}$) with the energy separation ~130 meV, which is the same as the gap-like feature generated by the *K* and *K'* out-of-plane phonons in zero magnetic field. Simultaneously, there are several peaks, as marked with $P_N$ ($N \neq 0$), depending on the square-root of both level index *N* and magnetic field *B* (see Fig. S3 of the Supplemental Material for details of analysis [33]), which are similar to that of the $LL_N$ for massless Dirac fermions in graphene monolayer with only a shift of ~ 65 meV in energy [35,36]. A close examination of the low-bias spectra also reveals the existence of the zero LL ($LL_0$) around the Fermi level, as shown in Fig. 1(e) (see Fig. S4 for more spectra [33]). All the experimental results indicate that the *K* and *K'* out-of-plane phonons play a vital role in the emergence of the new discrete states of graphene in magnetic fields.

In the absence of magnetic field, the *K* and *K'* out-of-plane phonons mix the nearly free-electron bands at Γ and the linear π bands of graphene and create new inelastic tunneling channels in the perpendicular direction of graphene [19,37]. Since the Pauli's exclusion principle blocks electrons tunneling into occupied states, there exist a threshold energy $\hbar\omega_{ph}$ determined by the vibrational frequency $\omega_{ph}$, below which the inelastic scattering process cannot happen (in our experiment, $\hbar\omega_{ph}$ ~ 65 meV is the energy of the *K* and *K'* out-of-plane phonons). Therefore, we can detect a gap-like feature with the energy ~ $2\hbar\omega_{ph}$ in the tunneling spectra of graphene monolayer, as schematically shown in Fig. 2(a) and experimentally observed in Fig. 1(b). By applying a perpendicular magnetic field, the massless Dirac fermions in graphene monolayer are condensed into discrete LLs. Then the coupling of phonons and quasiparticles in the LLs is expected to introduce a new set of peaks at energies $E_N + \hbar\omega_{ph}$ for $E_N > E_F$ and $E_N - \hbar\omega_{ph}$ for $E_N < E_F$, corresponding to the emission or absorption of phonons in the tunneling process, as schematically shown in Fig. 2(b). Therefore, the observed peaks in Fig. 1(c) and Fig. 1(d) should be attributed to the quantized phonon modes of the *K* and *K'* out-of-plane phonons in graphene. Previously, effects of coupling between quasiparticles in the LLs and a $E_{2g}$ phonon in graphene were calculated in theory and the quantized phonon modes of the $E_{2g}$ phonon were predicted [38,39]. However, the signal of the quantized phonon modes of the $E_{2g}$ phonon is predicted to be much weaker than that of the LLs and the main feature that predicted to be observed in experiment is the splitting of a LL in case of sufficiently closing to a quantized phonon peak [38,39]. Our experiment shows that this is not the case for the *K* and *K'* out-of-plane phonons: the intensity of the quantized phonon peaks is about 50 times compared with that of the LLs in the spectra (Fig. 1).

To further understand the observed giant quantized phonon mode, we calculate electron-phonon self-energy with considering the coupling of quasiparticles in LLs to the *K* and *K'* out-of-plane phonon with the energy of 65 meV [38]. Figure 2(c) shows the obtained quantized phonon modes as a function of magnetic fields. Two peaks $P_{0+}$ and $P_{0-}$, which are generated from the coupling of the $LL_0$ and the $K$ and $K'$ out-of-plane phonon, are independent of magnetic field. The other quantized phonon modes $P_N$ (with $N \neq 0$) are generated from the $LL_N$, depending on the square-root of both level index $N$ and magnetic field $B$ (Tab. S1 of the Supplemental Material [33]). Obviously, our simulation reproduces quite well the main features observed in our experiment (Fig. 1(d)). The only experimental feature that cannot be reproduced in our simulation is that all the quantized phonon peaks split into two peaks with the energy separation of about 8 meV, which is invariable in different magnetic fields (see Fig. 1(c), Fig. 1(d) and Fig. S5 of the Supplemental Material [33]). By considering the fine structure in the neighborhood of the phonon-emission threshold, these splittings are most likely attributed to the formation of bound states between an electron and an optical phonon [40]. Theoretically, the energy separation of the splitting in a quantized phonon peak induced by the bound states can be estimated as $\alpha\omega_{ph}$, where $\alpha$, the electron-phonon coupling parameter, is estimated to range from 0.04 to 0.13 in graphene [40]. In our experiment, $\alpha \sim 0.12$ is deduced according to the observed splitting.

Previously, it has been demonstrated explicitly that the phonon-mediated tunneling spectra of graphene can be controlled by the interaction between graphene and substrates [22-27,41-43]. In our sample, the interaction between graphene and the supporting substrate varies spatially due to the existence of nanoscale ripples or wrinkles. Figure 3(a) shows a representative STM image of graphene monolayer with a one-dimensional wrinkle. The tunneling spectrum acquired on the wrinkle exhibits a 130 meV gap-like feature pinned to the Fermi energy (red dots in Fig. 3(b)), implying the existence of the $K$ and $K'$ out-of-plane phonon excitations. However, the $dI/dV$ spectrum acquired on the flat graphene region exhibits the V shape (blue dots in Fig. 3(b)), which directly reflects local DOS of massless Dirac fermions in graphene. Similar observations are obtained around many different graphene wrinkles with different STM tips. Such a result indicates that the interaction between graphene and substrates can sufficiently suppress the $K$ and $K'$ out-of-plane phonons of graphene and, consequently, weaken the inelastic tunneling channels. Our experiment in high magnetic fields, as shown subsequently, further demonstrates the possibilities to switch on/off the quantized phonon modes in the tunneling spectra by changing the interactions between graphene and the substrate. Figure 3(c) shows the spatial-resolved *dI/dV* spectra recorded across the graphene wrinkle (the dotted arrow in Fig. 3(a)) by applying a magnetic field of 10 T (*dI/dV* spectra measured in different magnetic fields are shown in Fig. S6 and Fig. S7 of the Supplemental Material [33]). High-quality well-defined LLs of the massless Dirac fermions in graphene monolayer are obtained off the wrinkle, as shown in the right part of Fig. 3(c), indicating that the *dI/dV* spectra on the flat region mainly reflect the local DOS of graphene. With approaching the top of the graphene wrinkle, two pronounced peaks ($P_{0+}$ and $P_{0-}$), which are symmetry about the Dirac point ($LL_0$), appear and gradually increase in intensity. Simultaneously, the intensities of the $LL_0$ and $LL_{\pm 1}$ decrease with approaching the wrinkle. On top of the graphene wrinkle, the signal of the quantized phonon modes is much stronger than that of the LLs in the spectra. Such a result is further verified by the atomic-scale *dI/dV* maps measured around the wrinkle (Fig. S8 of the Supplement Materials [33]). Therefore, our experiment demonstrates explicitly the switch on/off of the quantized phonon modes at nanoscale by the interactions between graphene and the substrate.

Theoretically, the quantized phonon modes can be tuned by changing doping of graphene [38,39]. As schematically shown in Fig. 4(a), when the $LL_0$ is fully occupied, the $LL_0$ associated phonon mode only appear on one side of the Fermi level due to the Pauli exclusion principle (see Tab. S1 of the Supplemental Material [33]). In order to verify this doping-dependent phenomenon, we carry out similar high-field measurement in a graphene region that is electron doped. Figure 4(b) shows a representative zero-field STS spectrum of the studied graphene region, with the Dirac point well below the Fermi level due to the charge transfer between the topmost graphene layer and substrate. The Dirac point of the graphene monolayer is measured at about −70 meV according to the $LL_0$ measured in high magnetic field (Fig. S9 of the Supplemental Material [33]). Figure 4(c) shows a typical *dI/dV* spectrum recorded in this region in a magnetic field of 10 T and only one quantized phonon peak associated to the $LL_0$ is observed, as marked by $P_0$. Our simulation by taking into account the position of the $LL_0$, as shown in Fig. 4(d), reproduces well the experimental result.

Our experiment further demonstrates that it is possible to tune the quantized phonon modes by taking advantage of the STM tip pulse. By using a voltage pulse of 3 V for 0.1 s duration generated between the STM tip and the sample, we can locally enhance the van der Waals interactions between the topmost graphene sheet and the supporting substrate. Therefore, the $K$ and $K'$ out-of-plane phonon-mediated inelastic channel is suppressed and the quantized phonon modes become weak dramatically (see Fig. S10 of the Supplemental Material [33]). Moreover, our experiment demonstrates that the Fermi velocity in graphene monolayer increases about 10% when the $K$ and $K'$ out-of-plane phonon is suppressed. Such a result agrees with the theoretical calculations that the electron-phonon interaction can reduce the Fermi velocity of the massless Dirac fermions to $v_F^{ep} = \frac{v_F^0}{1+\alpha}$, where $v_F^0$ is the Fermi velocity when the electron-phonon interaction is negligible [44,45]. According to the measured Fermi velocities, $\alpha$ is estimated as about 0.10, consistent with that, ~ 0.12, deduced from the electron-phonon bound states.

In summary, we systematically study the electron-phonon interaction in graphene monolayer in high magnetic fields and observe giant quantized phonon modes of the $K$ and $K'$ out-of-plane phonons for the first time. Our experiments also show the ability to tune the giant quantized phonon modes at nanoscale by changing interactions between graphene and substrate. The weak van der Waals interactions between graphene and the substrate can be used as an effective "switch" to turn on or off the quantized phonon modes in graphene.

**Acknowledgements**

This work was supported by the National Natural Science Foundation of China (Grant Nos. 11674029, 11422430, 11374035), the National Basic Research Program of China (Grants Nos. 2014CB920903, 2013CBA01603). L.H. also acknowledges support from the National Program for Support of Top-notch Young Professionals, support from "the

Fundamental Research Funds for the Central Universities", and support from "Chang Jiang Scholars Program".

# Figures

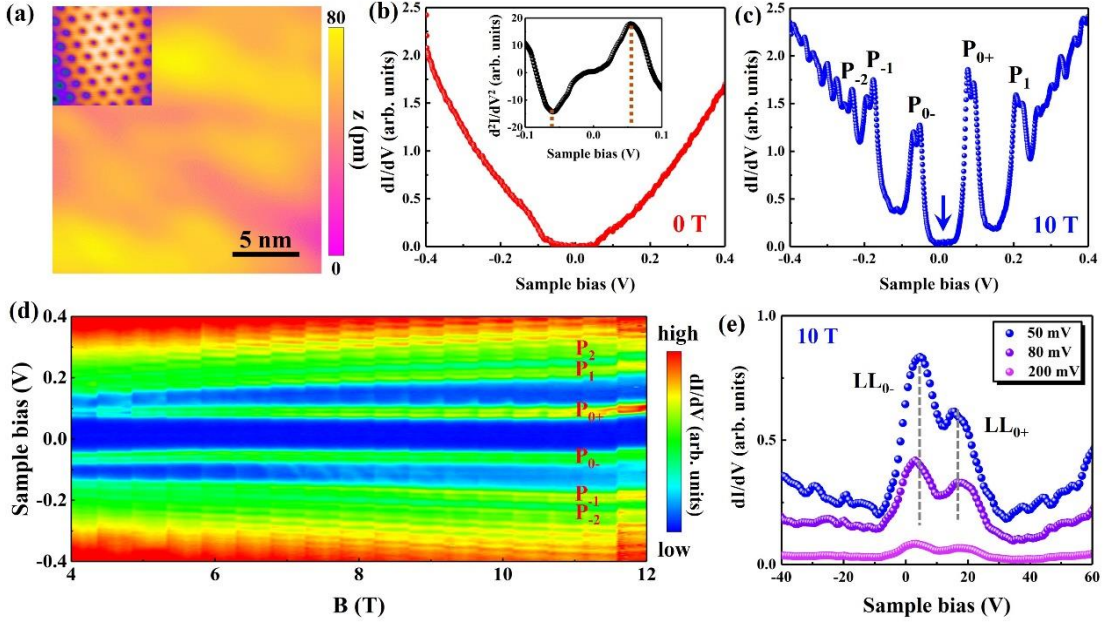

**Fig. 1.** The inelastic tunnelling spectra of suspended graphene monolayer under different magnetic fields. (a) A 20 ×20 nm$^2$ STM image of a suspended graphene region ($V_b$ = 0.3 mV, $I$ = 0.1 nA). Inset: Zoom-in atomic-resolution STM topography. (b) The average of a series of *dI/dV* spectra acquired in panel (a) at *B* = 0 T. The central gap-like feature is attributed to the *K* and *K'* out-of-plane phonons with the energy $\hbar\omega_{ph} \approx$ 65 meV in suspended graphene monolayer. Inset: *d$^2$I/dV$^2$* spectra acquired in panel (a) at *B* = 0 T. (c) The average of a series of *dI/dV* spectra acquired in panel (a) at *B* = 10 T. $P_N$ label the phonon peaks, which are induced by the intrinsic Landau levels coupling to the *K* and *K'* out-of-plane phonons. (d) The evolution of a series of inelastic electron tunnelling spectra as a function of magnetic fields *B* from 4 to 12 T as a variation of 0.5 T. Each panel of every *B* consists of thirty *dI/dV* spectra acquired at different spatial points along a line of 20 nm. (e) Enlarged *dI/dV* spectra within the gap marked by blue arrow in panel (c) measured with the different $V_b$ and in the magnetic field of *B* = 10 T (*I* = 0.4 nA). The $LL_0$ can be clearly seen inside the gap under low $V_b$.

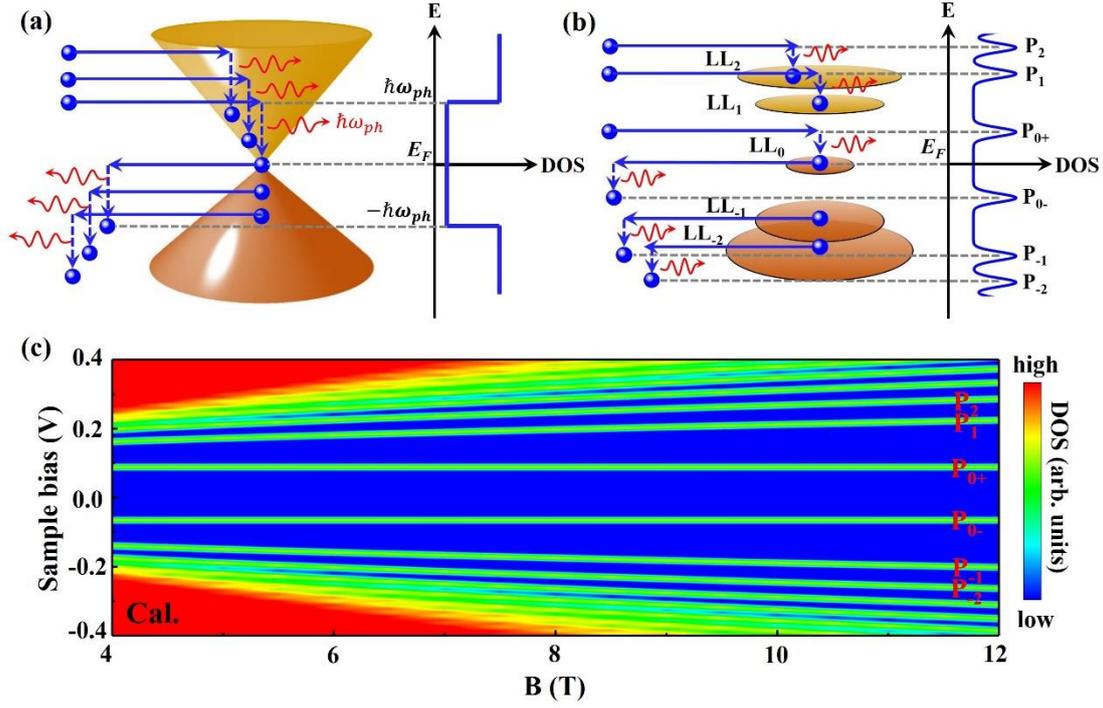

**Fig. 2.** Schematic images of the inelastic electron tunnelling spectra in graphene monolayer under different magnetic fields. (a) Schematic of the inelastic electron tunnelling due to excitation of phonons with the energy $\hbar\omega_{ph}$ at $B = 0$ T. The low-energy electronic structures for graphene show the two Dirac cones that meet at the Dirac point ($E_D$), where the density of carriers vanishes. The tunnel current is abruptly enhanced if the energy is high enough to excite a phonon with the threshold energy of $\hbar\omega_{ph}$, thus opening a new inelastic tunnelling channel at energies of $\pm\hbar\omega_{ph}$ and causing steps in DOS symmetric to $E_F$. (b) Schematic of the inelastic electron tunnelling due to excitation of phonons with energy $\hbar\omega_{ph}$ at $B > 0$ T. The intrinsic landau levels are shown as the numbered disks. (c) Calculations of the *B*-dependent density of states of the suspended graphene monolayer by considering the coupling of electron LLs and the *K* and *K'* out-of-plane phonons.

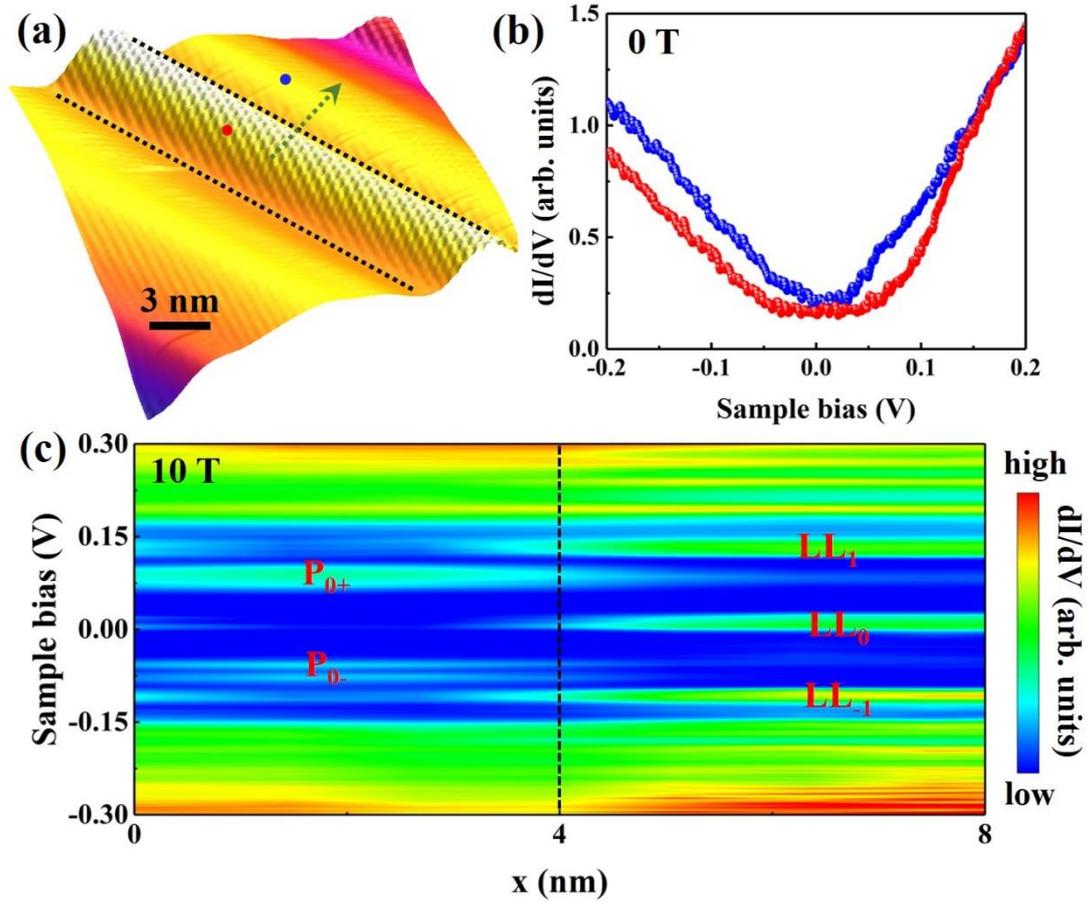

**Fig. 3.** Spatial variation of the *dI/dV* spectra around a graphene wrinkle. (a) A 20 × 20 nm² STM image of the typical graphene wrinkle in 3D view ($V_b$ = 0.3 mV, $I$ = 0.1 nA). The dashed lines indicate the edges of the wrinkle. (b) Typical *dI/dV* spectra at $B$ = 0 T recorded at different positions in panel (a). The red and blue dots indicate the positions where we acquired the corresponding *dI/dV* spectra. (c) The evolution of the *dI/dV* spectra acquired at different spatial points along the green arrow of panel (a) in the magnetic field of $B$ = 10 T. $P_N$ and $LL_N$ label the quantized phonon peaks and LL peaks, respectively.

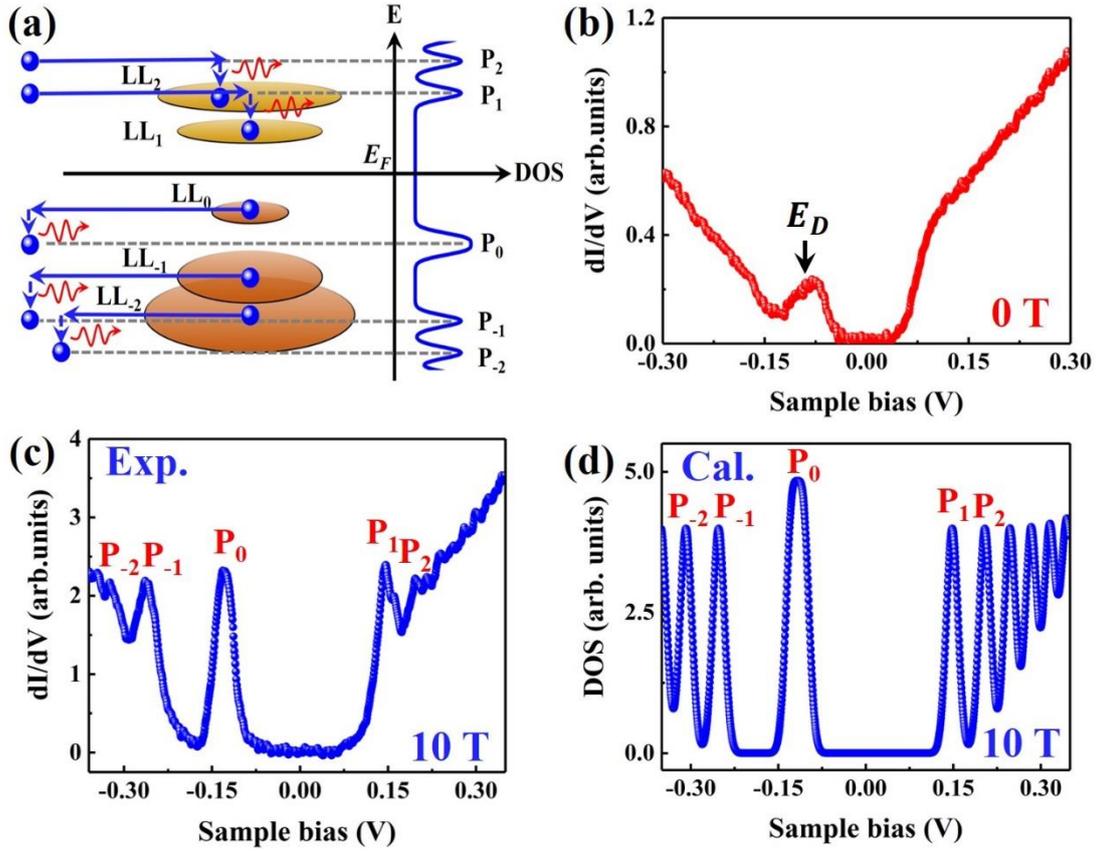

**Fig. 4.** The *dI/dV* spectra recorded on the graphene wrinkle with a large charge transfer. (a) Schematic of inelastic electron tunnelling due to excitation of phonons with energy $\hbar\omega_{ph}$ at $B > 0$ T in a hole-doped region. The intrinsic landau levels of graphene are shown as the numbered disks. (b) A typical *dI/dV* spectrum recorded on the graphene wrinkle at $B = 0$ T. The Dirac point, which is determined according to the zero Landau level in high magnetic field, is marked by the black arrow. (c) A typical *dI/dV* spectrum acquired on the graphene wrinkle at $B = 10$ T. $P_N$ label the phonon peaks. (d) Calculations of the density of states at $B = 10$ T by considering the coupling of the electron LLs and *K* and *K'* out-of-plane phonons in graphene.